\documentclass[11pt]{article}
\usepackage[margin=0.5in]{geometry}
\usepackage[numbers]{natbib}

\usepackage{amsmath, amssymb, graphicx}
\usepackage{caption}
\usepackage{tikz}
\usepackage{hyperref}

\newcommand{\ba}{\left( \begin{array}}
\newcommand{\ea}{\end{array} \right)}
\newcommand{\bq}{\begin{eqnarray*}}
\newcommand{\eq}{\end{eqnarray*}}
\newcommand{\bqn}{\begin{eqnarray}}
\newcommand{\eqn}{\end{eqnarray}}

\title{From Density to Void: Why Brain Networks Fail to Reveal Complex Higher-Order Structures}
\author{Moo K. Chung$^{*1}$, Anass B. El-Yaagoubi$^{2}$, Anqi Qiu$^{3}$, Hernando Ombao$^{2}$\\
$^{1}$Department of Biostatistics and Medical Informatics, University of Wisconsin, Madison, USA\\
$^{2}$Statistics Program, King Abdullah University of Science and Technology, Thuwal, Saudi Arabia\\
$^{3}$Department of Health Technology and Informatics, Hong Kong, China\\
\texttt{mkchung@wisc.edu}}
\date{}

\begin{document}
\maketitle
\thispagestyle{empty}

\begin{abstract}
In brain network analysis using resting-state fMRI, there is growing interest in modeling higher-order interactions beyond simple pairwise connectivity via persistent homology. Despite the promise of these advanced topological tools, robust and consistently observed higher-order interactions over time remain elusive. In this study, we investigate why conventional analyses often fail to reveal complex higher-order structures—such as interactions involving four or more nodes—and explore whether such interactions truly exist in functional brain networks. We utilize a simplicial complex framework often used in persistent homology to address this question.
\end{abstract}

\pagestyle{empty}

\section{Introduction}
\begin{figure}
\centering
\includegraphics[width=0.7\linewidth]{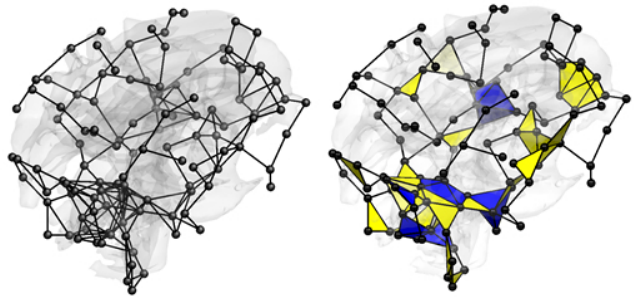}
 \caption{The traditional graph representation (left) and the proposed simplical complex representation (right) of a brain network. 2-simplices (yellow triangles) show the connectivity of three regions while 3-simplices (blue tetrahedrons) show the connectivity of four regions \citep{anand.2023.TMI}. Standard graph data structure cannot encode such complex connectivity patterns directly {\em without} additional models.}
 \label{fig:complex}
 \end{figure}

Recent advances in brain network analysis using resting-state fMRI have spurred interest in modeling interactions that go beyond conventional pairwise connectivity \cite{anand.2023.TMI,dakurah.2022,santoro.2024}. Traditional methods based on correlation matrices rely on a dyadic assumption, capturing only the interactions between two brain regions at a time. However, emerging evidence indicates that the functional architecture of the brain may involve complex, multiregional interactions that cannot be fully described by pairwise measures alone \cite{lucas.2020}. Moreover, conventional graph-based models, which represent connectivity as a network of nodes and edges, are inherently limited in encoding these higher-order relationships—such as three-way or four-way interactions—without additional analytical steps \cite{sporns.2003} (Figure~\ref{fig:complex}).

There have been several attempts to fit a single model to a selected number of $k$-node subsets for modeling k-node connectivity; however, this approach is problematic. First, the choice of which subsets to include is often arbitrary, potentially introducing selection bias and failing to capture the full diversity of higher-order interactions within the network. The significance of such a selection must be evaluated in the context of the entire sample space of $\binom{p}{k}$ possible combinations. Second, the combinatorial explosion—where the number of possible subsets grows exponentially—renders it impossible for a single model to represent the vast heterogeneity among these combinations. Third, such models are often highly complex, and fitting a model over the entire sample space is computationally prohibitive.

To address these limitations, several recent studies have attempted to model higher-order connectivity in brain networks using simplicial complexes \citep{anand.2023.TMI,lucas.2020,santoro.2024}. This approach overcomes by constructing a simplicial complex from the connectivity matrix. In this framework, nodes (0-simplices) represent individual time series, and an edge (1-simplex) is included when the connectivity between two regions exceeds a tunable threshold. Similarly, a triangle (2-simplex) is formed if all three pairwise connections among a triplet of regions exceed the threshold; higher-order simplices are defined analogously \citep{chen.2011, edelsbrunner.2000, lee.2014.MICCAI}. Persistent homology, a cornerstone of topological data analysis (TDA), then enables us to quantify multiscale topological features—such as connected components (0D features), cycles (1D features), and higher-dimensional voids—in these simplicial complexes. This framework offers a clear and tractable solution for revealing higher-order interactions in brain networks \citep{chung.2019.ISBI, chung.2019.NN, huang.2020.NM}.

Despite the promise of advanced topological tools, robust and consistently observed higher-order interactions over time remain elusive \citep{anand.2023.TMI, lucas.2020, santoro.2024}. Numerous studies have applied persistent homology to brain networks \citep{bourakna.2023.entropy, bourakna.2024, chung.2023.NI}, yet conventional analyses often produce networks that are either overly dense or highly fragmented, thereby obscuring the emergence of meaningful higher-order topological features. Moreover, as the number of nodes involved in an interaction increases, the probability that multiple regions activate simultaneously decays exponentially, further limiting the detection of robust higher-order structures.

Compounding these challenges, existing studies utilizing simplicial complexes frequently suffer from significant statistical shortcomings \citep{lucas.2020, santoro.2024}. In a network with \(p\) nodes, there are \(\binom{p}{k}\) possible subsets of \(k\) nodes, a number that grows exponentially with \(k\). This combinatorial explosion gives rise to a severe multiple comparisons problem \cite{benjamini.1995,benjamini.2001}. For instance, if the significance level for each \(k\)-node connectivity test is set at 0.05, then 5\% of these tests could be falsely declared significant purely by chance. Given the enormous number of potential interactions, this can lead to a large number of false positives. To date, none of the existing studies have adequately addressed this issue, underscoring the need for rigorous statistical methods to control the false discovery rate when assessing higher-order interactions.

%

In this study, we investigate why conventional connectivity analyses often fail to reveal complex higher-order structures—such as interactions involving four, five, or more nodes—and explore whether these interactions truly exist in functional brain networks. Rather than fitting a single model to an arbitrarily selected subset of \(k\)-node combinations—a strategy that risks selection bias and overlooks the vast heterogeneity inherent in the \(\binom{p}{k}\) possible interactions \citep{lucas.2020}—we propose to determine the existence of \(k\)-node interactions through simplicial complexes. Our approach explicitly acknowledges the combinatorial explosion and the associated multiple comparisons problem. To address these challenges, our framework incorporates rigorous false discovery rate correction methods that are applied across the entire sample space of interactions, thereby enabling robust statistical inference for detecting true higher-order connectivity patterns in the brain.

\section{Methods}

\subsection{Simplicial Complexes as Higher-Order Interactions}

Consider a multivariate time series \(X(t) = [X_1(t), X_2(t), \dots, X_p(t)]\), where each \(X_i(t)\) represents the activity of a brain region. We first compute a pairwise similarity measure, typically the Pearson correlation,
\[
C_{ij} = \text{corr}(X_i(t), X_j(t)),
\]
which yields a weighted connectivity matrix \(C\) that captures the strength of interaction between each pair of nodes. Other similarity or distance measures, such as entropy or mutual information \cite{maes.1997,nicolini.2020}, can also be employed; however, these will not be considered in the present study. To capture interactions beyond pairwise relationships, we represent the connectivity matrix \(C\) as a simplicial complex. In this construction, each node (0-simplex) corresponds to an individual time series, and an edge (1-simplex) is included when the connectivity between two regions exceeds a chosen threshold \(\tau\). Likewise, a triangle (2-simplex) is formed if all three pairwise connections among three regions exceed \(\tau\), and higher-order simplices are defined in a similar fashion. 

In the proposed hierarchical construction, the simplicial complex \(S\) has the property that if a \((k+1)\)-simplex \(\sigma^{k+1}\) is included in \(S\), then every \(k\)-dimensional face of \(\sigma^{k+1}\) is automatically included in the \(k\)-skeleton \(S^{(k)}\). 
That is, for each vertex \(v \in \sigma^{k+1}\), the face
\[
\sigma^{k+1} \setminus \{v\}
\]
is an element of \(S^{(k)}\). This hierarchical scheme ensures that all higher-dimensional simplices are built upon the complete set of their lower-dimensional faces. This combinatorial structure, which converts the connectivity matrix into a format suitable for persistent homology, enables the systematic exploration of higher-order interactions.

\subsection{Construction of the Boundary Matrices}
While the simplicial complex provides a structural representation by including nodes, edges, triangles, and higher-order simplices based on connectivity thresholds, it does not inherently quantify the significance of these interactions. The transition from simplicial complexes to boundary matrices is crucial for quantifying higher-order interactions in multivariate time series. To systematically measure higher-order interactions, we construct boundary matrices \(\partial_k\) that capture the relationships between \(k\)-simplices and their \((k-1)\)-dimensional faces. Each boundary matrix \( \partial_k \) represents the boundary operator acting on \( k \)-simplices, and it encodes how these \( k \)-dimensional simplices are built from their \((k-1)\)-dimensional faces \citep{anand.2023.TMI,huang.2023}. In other words, the boundary operator maps each \( k \)-simplex to a formal sum of its \((k-1)\)-simplices.

\begin{figure}
\begin{center}
\begin{tikzpicture}[scale=1.5, 
    every node/.style={draw, circle, minimum size=0.8cm}, 
    every path/.style={thick}]

    \node (v1) at (0,2) {$v_1$};
    \node (v2) at (2,2) {$v_2$};
    \node (v3) at (2,0) {$v_3$};
    \node (v4) at (4,0) {$v_4$};

    \draw[->] (v1) -- (v2)
      node[midway, above, draw=none, fill=none] {\( e_{12} \)};

    \draw[->] (v1) -- (v3)
      node[midway, above, xshift=-22pt, draw=none, fill=none] {\( e_{13} \)};

    \draw[->] (v2) -- (v3)
      node[midway, above, xshift=9pt, draw=none, fill=none] {\( e_{23} \)};

    \draw[->] (v2) -- (v4)
      node[midway, above, xshift=3pt, draw=none, fill=none] {\( e_{24} \)};
\end{tikzpicture}
\end{center}
 \caption{An example of simplicial complex consisting of 4 nodes.}
 \label{fig:4nodes}
\end{figure}
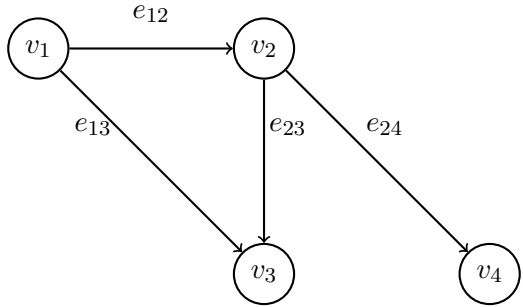

The 1-boundary matrix encodes the incidence and orientation of each edge relative to its incident nodes (0-simplices). For example, consider the directed graph in Figure~\ref{fig:4nodes}, where edge \(e_{12}\) connects node \(v_1\) to node \(v_2\). The boundary operator for this edge is given by
\[
\partial_1(e_{12}) = v_2 - v_1.
\]
In the matrix representation of \(\partial_1\), each column corresponds to a 1-simplex (edge) and each row corresponds to a 0-simplex (node). Thus, in the column corresponding to \(e_{12}\), the entry in the row for \(v_1\) is \(-1\) and the entry in the row for \(v_2\) is \(+1\), with zeros elsewhere. 1-boundary matrix corresponding to Figure~\ref{fig:4nodes} is given by
\[
\partial_1 =
\begin{bmatrix}
-1 & 1 & 0 & 0 \\
-1 & 0 & 1 & 0 \\
0 & -1 & 1 & 0 \\
0 & -1 & 0 & 1
\end{bmatrix}.
\]
This matrix succinctly captures the connectivity and orientation information for the edges in the graph. Similarly, the 2-boundary matrix \( \partial_2 \) describes the relationship between triangles (2-simplices) and edges (1-simplices), where each row represents a triangle, and each column represents an edge. 

For any \( k \)-simplex \( \sigma^k = [v_0, v_1, \dots, v_k] \), its boundary consists of \( (k-1) \)-dimensional faces. The boundary matrix \( \partial_k \) is computed similarly as 
\[
\partial_k (\sigma^k) = \sum_{i=0}^{k} (-1)^i [v_0, v_1, \dots, \hat{v}_i, \dots, v_k],
\]
where \( \hat{v}_i \) denotes that the vertex \( v_i \) is removed. The corresponding boundary matrix is then given by
\[
(\partial_k)_{ij} =
\begin{cases}
+1, & \text{if } \sigma_j^{k-1} \text{ is a face of } \sigma_i^k \text{ with aligned orientation}, \\
-1, & \text{if } \sigma_j^{k-1} \text{ is a face of } \sigma_i^k \text{ with opposite orientation}, \\
0, & \text{otherwise}.
\end{cases}
\]

\subsection{Computational Complexity}

A brute-force approach that computes each \(\partial_k\) independently can be extremely inefficient and can cause a serious computational bottleneck for large $k$. Instead, we  will build \(\partial_{k+1}\) iteratively from \(\partial_k\) using the Intersecting Neighbor Sets algorithm \cite{schank.2005}. For example, going from \(\partial_1\) (edges) to \(\partial_2\) (triangles) requires finding all node triplets that form triangles. In a dense graph with \(p\) nodes and \(\mathcal{O}(p^2)\) edges, there can be \(\mathcal{O}(p^3)\) such triplets, making the naive enumeration \(\mathcal{O}(p^3)\). However, for sparse graphs or simplicial complexes where \(E \ll p^2\), more efficient methods such as intersecting neighbor sets can achieve around \(\mathcal{O}(E^{3/2})\) \cite{schank.2005}. Once a triangle is identified, populating the corresponding row in \(\partial_2\) is straightforward and adds only linear overhead in the number of triangles.

When constructing \(\partial_{k+1}\) from \(\partial_k\), a similar principle can be appliied for higher dimensions. Suppose each \(k\)-simplex contains nodes that each have at most \(k\) neighbors \emph{within} that simplicial structure. In other words, for each node of a given \(k\)-simplex, there are at most \(k\) other nodes that share a simplex with it. Under this assumption, intersecting the neighbor sets of \(k\) nodes takes \(\mathcal{O}(k)\) time. If \(n_k\) is the number of \(k\)-simplices, the total run time to identify all \((k+1)\)-simplices is thus \(\mathcal{O}(n_k \cdot k)\). This is much smaller than the worst-case \(\mathcal{O}(p^k)\) complexity, and in practice remains computationally feasible for moderate values of \(k\), particularly when connectivity matrices are thresholded in the construction of simplicial complexes. 

We have implemented the Intersecting Neighbor Sets algorithm for iteratively constructing a simplicial complex from a given connectivity matrix as a MATLAB function:
\begin{verbatim}
function S = PH_connectivity2simplex(C, tau, k)
\end{verbatim}
This function takes as input the correlation matrix \texttt{C}, a threshold \texttt{tau}, and the maximum dimension \texttt{k} for the simplicial complex. The function is part of the PH-STAT package, available at \url{https://github.com/laplcebeltrami/PH-STAT}.

\subsection{Checking for Higher-Order Interactions}

We compute the boundary matrices $\partial_1, \partial_2, \dots$ iteratively to capture the emergence and disappearance of topological features—such as loops (1-cycles) and voids (2-cycles)—that reveal complex network structures beyond simple pairwise connections. The mere presence of a triangle in a simplicial complex does not imply a meaningful higher-order interaction unless it contributes to a sustained topological feature. For a feature to be considered significant, it must persist over a range of threshold values \citep{carlsson.2009.bulletin,edelsbrunner.2002}. For example, if a triangle consistently contributes to forming a loop across various thresholds, this indicates that the underlying connectivity pattern is robust and reflects a genuine higher-order interaction; conversely, features that appear only briefly as the threshold varies are likely to be artifacts of noise.

To quantify the presence of higher-order interactions, we count the number of $k$-simplices observed in the constructed simplicial complex. In a network with $p$ nodes, there are $\binom{p}{k}$ possible $k$-node subsets. This combinatorial explosion gives rise to a severe multiple comparisons problem. If each potential $k$-node interaction is tested individually at a nominal significance level of 0.05, then 5\% of these tests could be falsely declared significant purely by chance \citep{benjamini.1995,benjamini.2001}. None of the existing studies have adequately addressed this issue, underscoring the need for more rigorous statistical methods to control the false discovery rate when assessing higher-order interactions.

We propose to quantify the relative frequency of higher-order interactions by computing the ratio
\[
\lambda_k =  \frac{N_k}{\binom{p}{k}},
\]
where $N_k$ is the observed number of $k$-simplices in the simplicial complex. A low value of $\lambda_k$ indicates that only a small fraction of the potential higher-order interactions is observed, suggesting a lack of robust higher-order connectivity. Conversely, a high value of $\lambda_k$ implies that a relatively larger fraction of the possible $k$-node interactions is present.

In many correlation-based brain networks, most node pairs exhibit significant correlation, resulting in a densely connected network when a low threshold is applied. This density minimizes the occurrence of gaps—missing connections that would otherwise form higher-dimensional features. Conversely, applying a high threshold may fragment the network, eliminating the structured cycles and voids necessary for capturing complex higher-order interactions. Thus, the inherent uniformity and density in these networks tend to suppress the emergence of distinct, nontrivial topological features. In contrast, methods such as Rips filtrations applied to point clouds naturally reveal connectivity at multiple scales and often expose rich higher-order structures. The thresholding process in correlation networks, however, typically yields networks that are either overly connected—masking potential gaps—or too sparse to support complex higher-order interactions. This discrepancy helps explain why robust higher-order topological features are rarely observed in conventional correlation network analyses, or with any connectivity measure that produces a dense connectivity matrix.

\section{Results}

\begin{figure}
\centering
\includegraphics[width=1\linewidth]{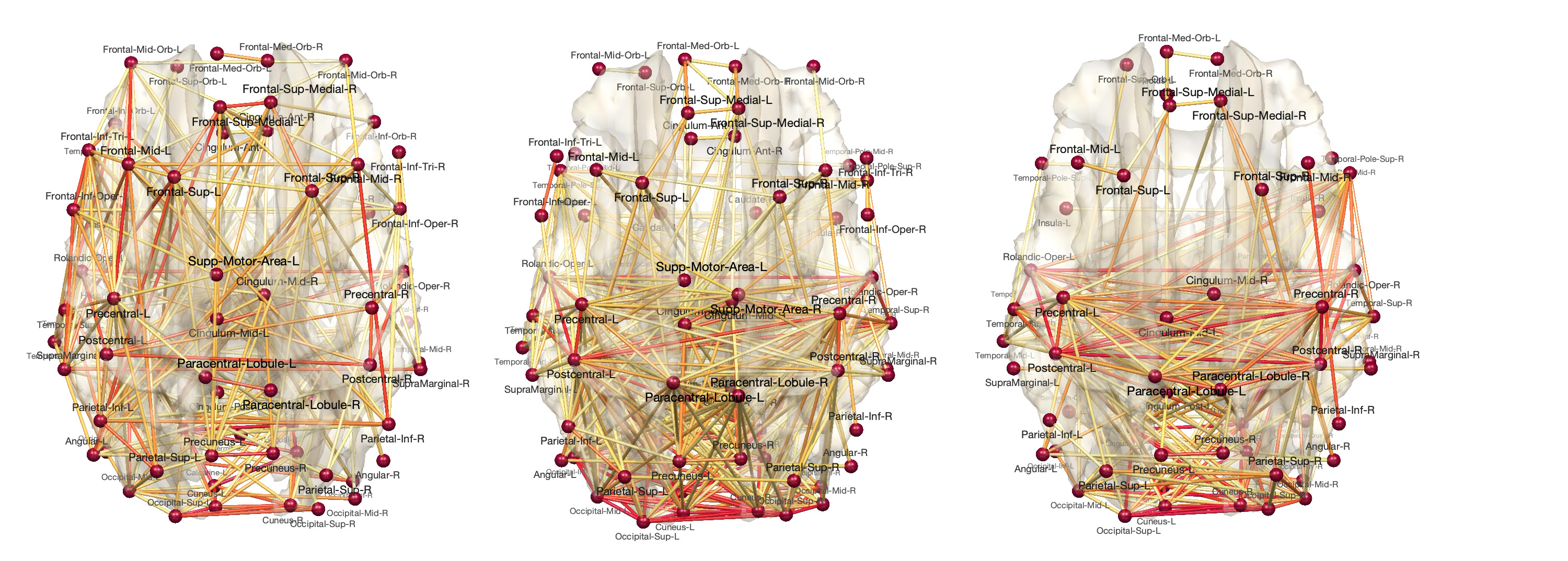}
\caption{Three representative rs-fMRI correlation brain networks. Due to the high density of connections, the networks are thresholded at 0.7 for visualization. While higher-order simplices may be present in individual networks, only 1- and 2-simplices are consistently observed across subjects. In contrast, the probability of detecting common 3-simplices or higher in all 100 subjects is effectively nil.}
 \label{fig:ratio}
 \end{figure}
 
In this study, we applied the method to a subset of resting-state fMRI data  in the Human Connectome Project (HCP) \cite{vanessen.2013}. The 100 healthy participants were aged between 22 and 36 years, with an average age of 29.05 \(\pm\) 3.36 years. Each fMRI scan lasted approximately 15 minutes while subjects maintained a relaxed state with their eyes open, fixating on a bright cross-hair against a dark background. Data were acquired on a customized Siemens 3T Connectome Skyra scanner using a gradient-echo planar imaging (EPI) sequence with a multiband factor of 8, a repetition time (TR) of 720 ms, an echo time (TE) of 33.1 ms, and a flip angle of 52°. The acquisition matrix was 104 \(\times\) 90 (RO \(\times\) PE), with 72 slices and 2 mm isotropic voxels, yielding a total of 1200 time points per scan.

\begin{figure}
\centering
\includegraphics[width=1\linewidth]{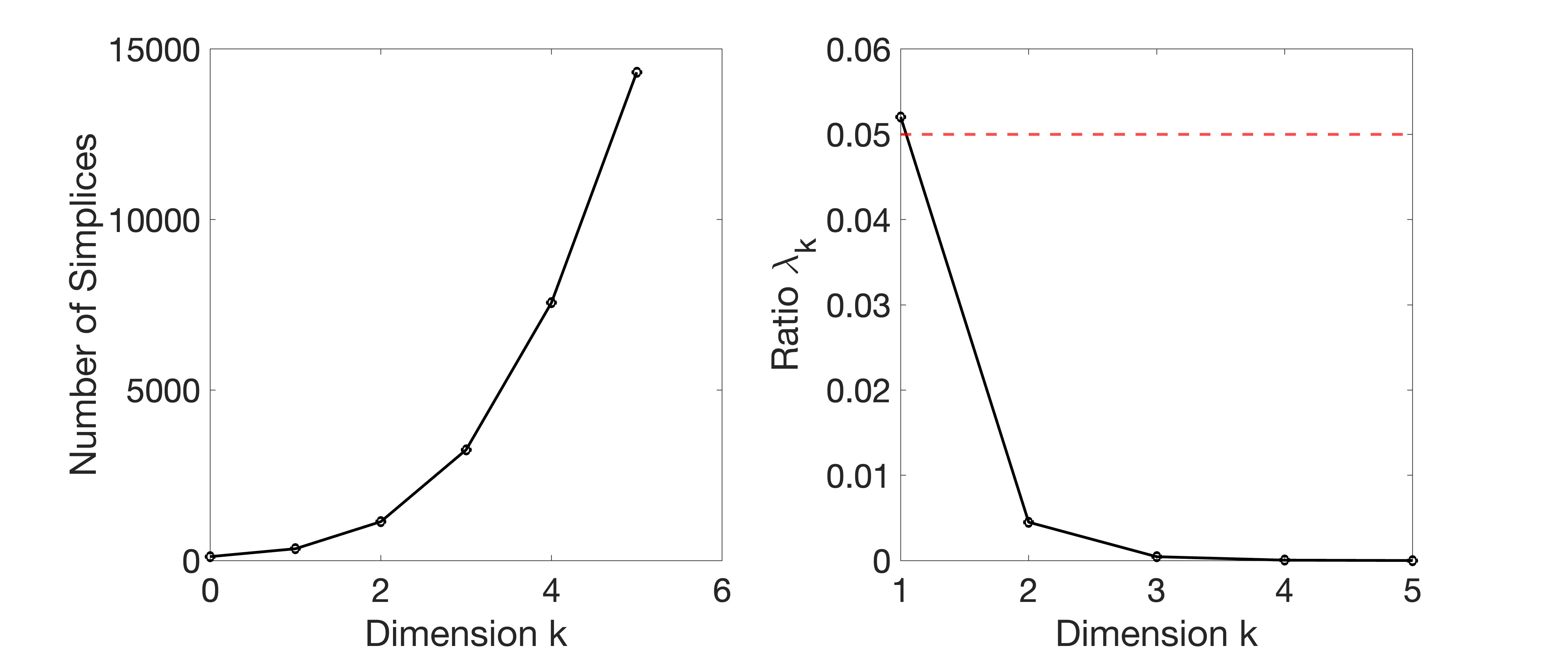}
\caption{Left: The average number of \(k\)-simplices observed across 100 subjects as a function of dimension. Right: The average ratio \(\lambda_k\) over dimension in 100 subjects. Since \(\lambda_k\) drops drastically after dimension 3, it is unlikely that higher-order interactions above 3-simplices are statistically significant. None of these simplices are likely to pass multiple comparisons correction at the 0.05 level.}
 \label{fig:ratio}
 \end{figure}
 
MRI preprocessing followed the HCP minimal preprocessing pipelines \citep{glasser.2013}, with additional standard procedures such as motion correction \citep{cox.1996}. A detailed account of our preprocessing pipeline is provided in our previous study \citep{huang.2020.NM}. After preprocessing, the resting-state functional time series were obtained in a volumetric space of \(91 \times 109 \times 91\) with 2 mm isotropic voxels over 1200 time points. Brain parcellation was performed using the Automated Anatomical Labeling (AAL) atlas \citep{tzourio.2002}, segmenting the brain into 116 distinct anatomical regions. The fMRI signal within each parcellated region was averaged across voxels, yielding 116 time series of 1200 time points per subject. Pearson correlations were then computed over the whole time point to generate static correlation brain networks \citep{shirer.2012, allen.2014, leonardi.2015}.

The 100 correlation matrices of size 116 $\times$ 116 were then used as input to the proposed methods in determining ratio $\lambda_k$ for various $k$.  We used $\tau=0.7$ as the threshold. Figure~\ref{fig:ratio} displays the average number of \(k\)-simplices observed across brain networks from 100 subjects. As the dimension increases, the total number of simplices grows exponentially, while the average ratio of observed to potential \(k\)-simplices decreases exponentially. For example, we observed \(\lambda_3 = 4.54 \times 10^{-4}\), \(\lambda_4 = 4.71 \times 10^{-5}\), and \(\lambda_5 = 4.83 \times 10^{-6}\). These ratios are far below a nominal significance level of 0.05, suggesting that higher-order simplices beyond 3-simplices are unlikely to represent robust higher-order interactions.

\begin{figure}[t]
\centering
\includegraphics[width=1\linewidth]{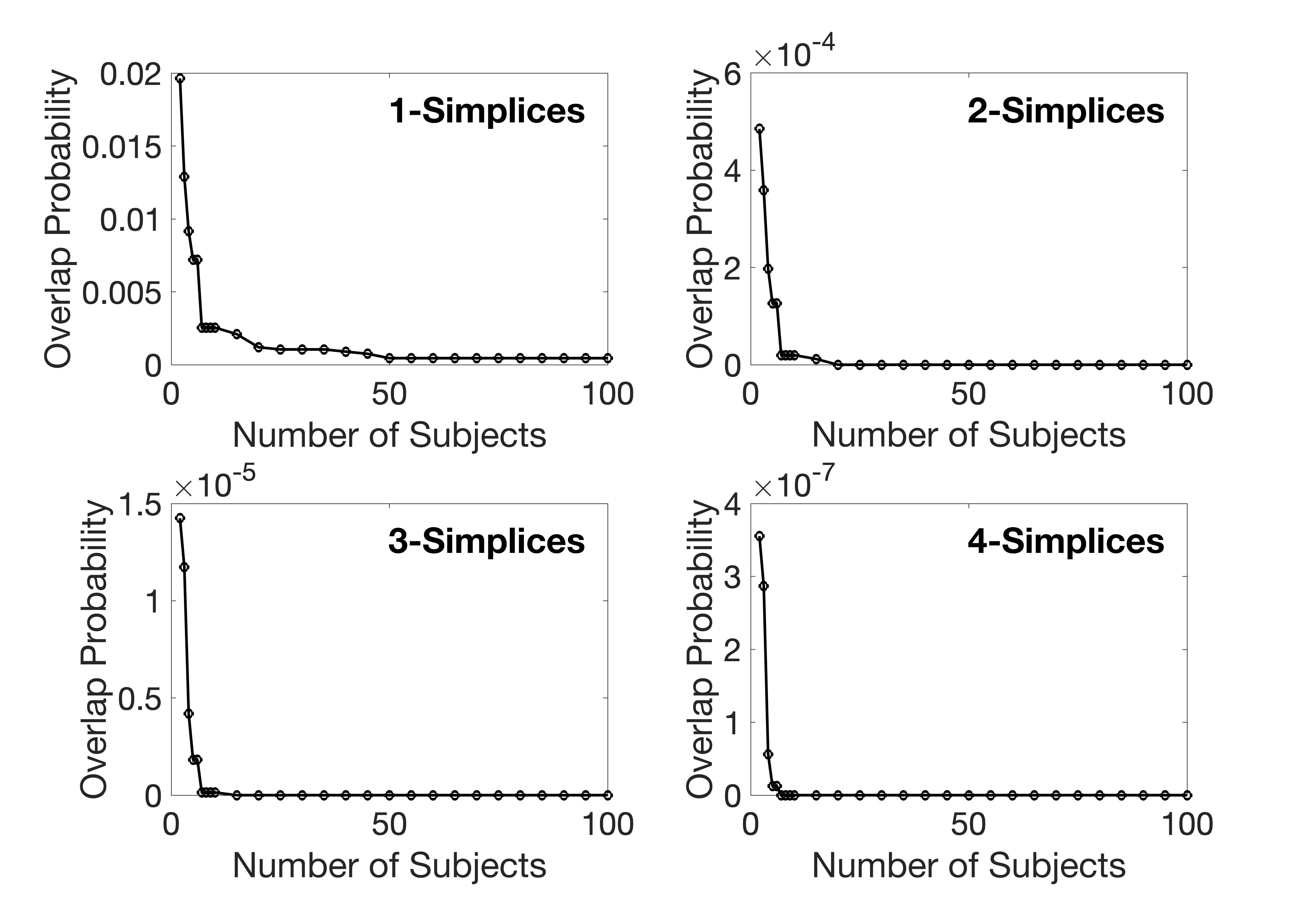}
\caption{Overlap probability of observing \(k\)-simplices across subjects. The plot shows the likelihood that a \(k\)-simplex—a group of \(k+1\) nodes sharing a higher-order interaction—is consistently present across subjects. The overlap probability decays exponentially with the number of subjects. Our results indicate that 1-simplices remain robust across 100 subjects, 2-simplices are detectable only up to about 20 subjects, and higher-order simplices vanish beyond 10 subjects, suggesting that robust higher-order interactions are extremely rare.}
 \label{fig:overlap}
 \end{figure}

We further investigated whether $k$-simplices are consistently observed across subjects by computing the overlap probability as the number of subjects increases (Figure~\ref{fig:overlap}). Since the occurrence of a particular $k$-simplex in an individual’s network can be modeled as an independent Bernoulli event, the probability that this simplex is present in all $n$ subjects decreases exponentially with $n$. Our results show that 1-simplices maintain a nonzero overlap probability even at 100 subjects, whereas 2-simplices exhibit nonzero overlap only up to about 20 subjects. Beyond this, the overlap probability for higher-order simplices rapidly declines to zero; in particular, 3-simplices and higher are virtually absent beyond 10 subjects. These findings strongly suggest that robust higher-order interactions (beyond pairwise and, to a lesser extent, triplet interactions) are not consistently observable in our datasets.

\section{Conclusion \& Discussion}

In this work, we simplicial complex used in persistent homology to quantify higher-order interactions in brain networks derived from resting-state fMRI. Our analyses reveal that the functional connectivity observed in these networks is predominantly governed by pairwise interactions, with only sparse evidence for three-node interactions. In contrast, robust interactions involving four or more nodes were essentially absent. These findings suggest that the intrinsic functional architecture of the brain is largely determined by lower-dimensional topological structures, specifically the patterns of connected components and cycles \citep{anand.2023.TMI,chung.2023.NI}.

Even though individual subjects may exhibit higher-order interactions, these interactions are not consistently observed across subjects. In our data, higher-order simplices rarely overlap between subjects, rendering further investigation of their persistence impractical. Based on these findings, future work should concentrate on the persistence of lower-dimensional interactions, particularly those involving 1-simplices (edges) and 2-simplices (triangles) \citep{lee.2014.MICCAI}.

While traditional TDA approaches often conceptualize higher‐order interactions as simultaneous connectivity among four or more nodes \citep{lucas.2020,santoro.2024}, we propose that the true complexity of brain network interactions arises from more subtle, temporally evolving lower‐dimensional interactions. In our view, robust higher‐order dynamics are not driven by isolated, large‐scale multi‐node events but rather emerge from the persistent interplay of pairwise (2-node) and triplet (3-node) connections. Over time, these recurrent interactions can combine in a nonlinear manner, giving rise to emergent network properties that reflect higher‐order organization. This perspective underscores the need for analytical frameworks that capture both the temporal persistence and the cumulative impact of simpler interactions. This is left as a future study.

\section{Acknowledgements}
NIH grants EB028753, MH133614 and NSF grant MDS-201077. We would like to thank D. Vijay Anand of University College London for generating Figure 1 that was originally published in \citep{anand.2023.TMI}.

\bibliographystyle{plainnat}
\bibliography{reference.2025.03.17}

\end{document}